# Gated Fusion Network for SAO Filter and Inter Frame Prediction in Versatile Video Coding.

Shiba Kuanar, Dwarikanath Mahapatra,Vassilis Athitsos, K.R Rao, *Fellow, IEEE,*

***Abstract*—To achieve higher coding efficiency, Versatile Video Coding (VVC) includes several novel components, but at the expense of increasing decoder computational complexity. These technologies at a low bit rate often create contouring and ringing effects on the reconstructed frames and introduce various blocking artifacts at block boundaries. To suppress those visual artifacts, the VVC framework supports four post-processing filter operations. The interoperation of these filters introduces extra signaling bits and eventually becomes overhead at higher resolution video processing. In this paper, a novel deep learning-based model is proposed for sample adaptive offset (SAO) nonlinear filtering operation and substantiated the merits of intra-inter frame quality enhancement. We introduced a variable filter size multi-scale CNN (MSCNN) to improve the denoising operation and incorporated strided deconvolution for further computation improvement. We demonstrated that our deconvolution model can effectively be trained by leveraging the high-frequency edge features learned in a parallel fashion using feature fusion and residual learning. The simulation results demonstrate that the proposed method outperforms the baseline VVC method in BD-BR, BD-PSNR measurements and achieves an average of 3.762 % bit rate saving on the standard video test sequences.**

***Index Terms*—Deep Learning, Artifacts, Versatile Video Coding, SAO Filter, Deconvolution.**

## I. Introduction

With the emergence of fast digital media development, ultra-high definition and virtual reality video have become extremely popular. These explosive growths of video data consume lots of network bandwidth and pose a challenge in storing and transmitting visual data over the channel. Therefore, it is necessary to improve the video compression performance and efficiently transmit high-quality videos. To meet this demand, the JCT-VC expert team proposed the Versatile Video Coding (VVC) standard [2] and adopts block- based hybrid coding schemes. Most compression block-based algorithms and VVC rely on tiling the images into several sub- blocks, applying quantization, motion estimation, and compensation, sparse transform, context-based adaptive coding, and frame reconstruction. Compared with its predecessor HEVC standard [1], the upcoming video coding standard, Versatile Video Coding achieves up to 50% BD-rate reduction. These high compression rates introduce undesired artifacts. The artifacts decrease the visual quality and therefore affect various image processing tasks such as image super-resolution [7], contrast enhancement [5], and edge detection [11]. Subsequently, the compressed image regularly blurs due to the loss of high-frequency components. Despite the progress of several video and image compression techniques, an effective artifact reduction technique remains an active area of research.

Dr. V. Athitsos and Dr. K.R. Rao with the Department of Computer and Electrical Engineering, University of Texas Arlington, USA.
S. Kuanar, with the Department of Electrical Engineering University of Texas Arlington and Yale University. e- mail:shiba.kuanar@mavs.uta.edu.
D. Mahapatra with the Inception Institute of Artificial Intelligence, UAE.

The block-wised operation is widely used in various modern video coding standards and encodes the video frame contents. Though block-based coding has its own merits, it leads to quantization noises, blocking, and ringing artifacts in decoded images. The blocking artifacts in video compression mainly arise from two aspects. Firstly, after quantization and inverse quantization, the difference of pixel values between blocks is magnified. Secondly, in the motion estimation and compensation process, the selected reference blocks are usually copied from various positions of different reference frames. Frequently, these matching blocks are not accurate and lead to pixel value discontinuity. Subsequently, the blocking artifact at the prediction block unit boundary weakens the input image's subjective quality. Additionally, the transformation of different frame-blocks follows independent coding processes with several parameters. These parameters in subsequent downstream processing result in the discontinuity of residual signals. In addition to that, the quantization distortion of high-frequency co-efficient causes the ripple phenomena around the sharp borders and introduces ringing artifacts. These ripple non- linear phenomena induce poor user experience. Therefore, a solution to minimize the blocking artifacts plays a significant role in real-time practical applications.

Four built-in loop filters were proposed for the VVC standard to alleviate the coding artifacts in compressed video and delineated in in Fig. 1. These decoder in-loop filters include: a deblocking filter [6], sample adaptive offset filter [15], adaptive loop filter [13], and Luma Mapping with Chroma Scaling (LMCS) [32]. A deblocking filter was adopted to target the blocking artifacts and smooth the boundary pixels while the block's inner pixels remain unchanged. Subsequently, an SAO filter was incorporated to reduce the sample distortion by adding an adaptive offset to each sample. An offset, a positive or negative integer, is obtained for each category and is then added to each sample. Currently, the adopted SAO in VVC comprises four one-dimensional edge patterns into consideration and uses in edge offset type selection [35]. However, if one block contains multiple edges, the four 1-D edges offset-based approach will not be efficient enough to remove the artifacts in all directions. The alternative approach to this problem is to adaptively combine different one-dimensional edge patterns for several edge types in a particular region. Subsequently, an adaptive loop filtering (ALF) technique suggested reducing the coding errors both in output and reference samples. Inspired by the Wiener filter theory, ALF trains filter coefficients to minimize the mean square error between reconstructed and original frames and reduce the overhead. These in-loop filters are stacked to overcome different distortions and improve visual quality while saving coding bit-rate.

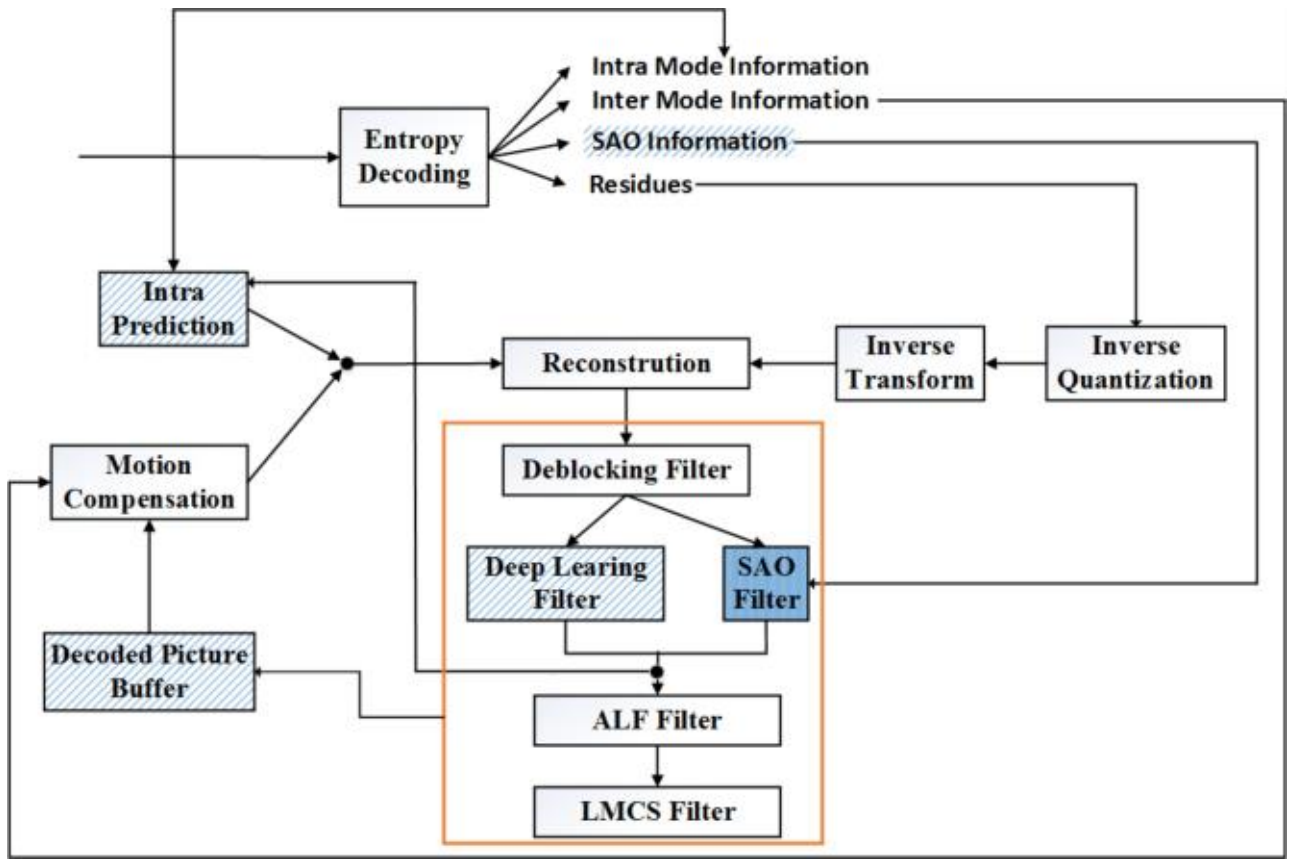

Fig. 1: VVC Decoder Diagram with SAO filter Replaced by our MSCNN Deep Learning Model.

Besides the built-in filters, a few alternative SAO filtering approaches have been suggested progressively in VVC. These techniques can be broadly classified either as a heuristic or a learning-based method. In heuristic methods [26], [20], the statistical properties of the artifacts are usually modeled according to the image priors. These statistical features include gradient, intensity, edges, colors, texture, shape, and varies within frames. Although these heuristic approaches boost the coding efficiency, the prior features are calculated manually in these methods. Thus, such hand-crafted feature extraction does not guarantee a good object descriptor and results in inefficiency in image quality improvement. Alternatively, it is intractable to build a multi-scale filtering framework and enhance the coding efficiency. To address the above shortcomings, learning-based methods are introduced to determine the broader image features and optimize with sufficient trainable parameters. Besides that, the artifact reduction methods can be extended barely from one compression method to another. Subsequently, data-driven designs can be an alternative for better generalization of image reconstruction and quality enhancement in compressed frames.

We introduced a deep learning-based method for artifact reduction in SAO output prediction. Our shallow MSCNN network includes the feature enhancement, de-noising, and extraction layers in succession and removes the undesired noisy features. Additionally, we also find difficulty while training the network layers due to sub-optimal initialization. To speed up the training process the model learned the residual blocks between the neighboring patches and used high learning rate values with an adjustable gradient clipping technique [3]. The residual blocks are calculated as the difference between the blocks in the current (target) frame to that of the blocks in the reference (matching) frame after the manipulation. We observed that the effective feature learning is improved by transferring the model parameters, fine-tuning the network variables, and applying data augmentation during the training. Finally, our model effectively suppressed the compressed artifacts to a great extent while retaining edge patterns, and sharp details. The result quantification shows that our MSCNN network outperformed previously studied networks in achieving better computational speed and higher bit rate reduction.

**Contribution**: We acknowledge the in-loop filtering as a design problem and it is difficult to dissociate the design of deep-learning approaches from that of the video decoder. Therefore, a systematic analysis is required to incorporate the CNN models in the video decoding framework. A new loss function is included for our model training. The global loss function measures the mean squared error for each batch and adds the intermediate spatial and temporal features for the final image reconstruction. In brief, the important contributions of our paper are summarized below.

- We introduced a gated fusion layer in our architecture design, which effectively combines the inter-frame local and temporal features. Our model includes a new loss function to constrain the pixel error and consists of an intermediate feature layer through skip connections. Our loss function measures the mean squared error at each batch and adds the residual sparse feature maps for image reconstruction. As a whole, our noise-robust loss function can be viewed as a generalization of MSE and low-level features learned from intermediate convolution layers.
- During the encoding process, different quantization parameters cause different levels of artifacts. To overcome it, a data-driven deconvolution model is integrated into decoder framework [8]. We collected a large-scale dataset for our SAO filter training. The end-to-end framework learns the feature in different sub-modules, optimizes the parameters, and removes the noise to a greater extent.
- Our model removes the artifacts in circumstances where targets are highly crowded and iteratively deconvolve the blurred images in a multi-layer framework. The result shows better image visibility with reduced blurring effects and did not suffer much from distortion.

The rest of the paper is organized as follows. Section II describes the related work and reviews the recent heuristic and learning-based approaches for VVC in-loop filter. Section III presents a summary of the proposed framework design. Section IV evaluates the experimental results and Section V analyses the result and it's trade-offs. Section VI reports the conclusion.

## II. Related Work

In this section, we briefly discussed the prior work related to the VVC loop filtering approach. The current in-loop filter usually resides in the video decoder loop, where the filtered blocks can be used as a prediction for later coded blocks or the reconstruction of the whole frame. Apparently, they change the rate-distortion behaviors of each CU, make inconsistency in the rate optimization and finally CTU level model decision at different QP values. Over the years, a couple of in-loop filtering techniques have been suggested to improve the coding efficiency and visual quality of the reconstructed frames in the VVC decoder. These techniques are applied to both intra and inter predictions and reconstruct the pixels before writing them into the decoded buffer [29]. In addition to the VVC framework development, four major built-in in-loop filters are proposed to reduce the compression artifacts: 1) DBF filter, 2) SAO filter, 3) ALF filter, and 4) Luma Mapping with Chroma Scaling (LMCS) operation. At the beginning of this subsection, we reviewed the existing in-loop filtering algorithms in VVC standard, following which the recent CNN-based image restoration method details are presented.

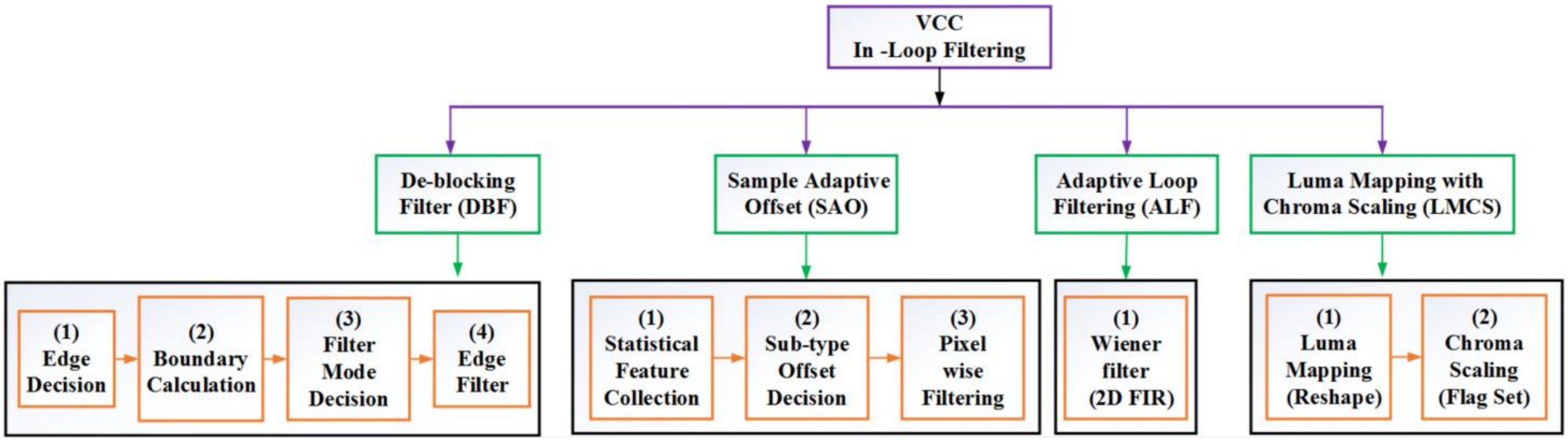

Fig. 2: Basic In-loop Filtering Steps in VVC Decoder and Encoder

### A. *In-Loop Filtering Methods in Video Coding*

The block-based transform coding method in different video coding standards introduces various artifacts and degrades the image reconstruction quality. Many post-processing methods are proposed in VVC to address these filtering problems, and several of them have already been adopted in the current standard. The primary motivation to design these low-pass filters is to confine the blocking artifacts by adaptively smoothing the boundary pixels and comparing adjacent boundary discontinuity with certain thresholds.

*1) Deblocking Filter:* The deblocking filter was initially introduced to remove the noise and blocking artifacts at the transform unit boundaries. As indicated in Fig. 2, the deblocking operation can split into four steps, namely 1) edge decision, 2) strength calculation at boundary regions, 3) filtering mode decision, and 4) edge filtering. The deblocking low pass filters in H.264 [15] and HEVC [19] originally applied to the 4×4 or 8×8 block boundaries. After that, it calculates the boundary strengths according to the encoder information and selects one of possible values 2, 1, and 0 indicating the strong, weak, and no filtering. If the boundary strength value is high, an additional condition check happens and applies weak or strong filtering mode based on sample values near the boundaries. Subsequently, edge filters are applied both horizontally and vertically for further smoothening.

*2) Sample Adaptive Offset:* Fu et al. introduced the SAO [17] in 2012 and later included in the HEVC Framework. The sample adaptive offset filter's design plan is to compensate reconstructed samples by adding an offset to each pixel and reducing the distortion between the frames. The SAO filter refines samples in both smoothed and irregular texture regions by dividing the samples into different sub-categories [27] and then adds offset to each sample value by encoder look-up tables. As pointed out in Fig. 2, SAO operation incorporates three major steps, precisely 1) statistical feature collection, 2) SAO sub-typing and offset decision, and 3) adaptive filtering. In step one, the number of pixels and their sum of distortions are calculated for a certain SAO type. In step two, the method's key task is to classify the reconstructed samples and select the offsets for each category. In the current implementation, the SAO filtering operation is categorized broadly into band offset (BO) and edge offset (EO) methods. After that, the best SAO type and its corresponding offset are selected from the candidates of edge offsets and band offset types. Overall, the edge offset reduces the artifacts around all edges directions, and the band offset reduces the artifacts in a relatively smooth region. Lastly, CTU pixels are filtered in step three by conditionally adding the corresponding best offset values.

*3) Adaptive Loop Filtering:* During the versatile video coding framework development, an advanced adaptive loop filtering (ALF) [13], [30], [31] was suggested and included after the SAO filter operation. The proposed ALF is located at the last processing stage of the in-loop process and can be regarded as a tool to capture and fix artifacts for each frame from previous settings. The ALF operation further minimizes the mean square error between all sample locations of reconstructed frames and the raw frames. ALF scans the decoded pictures at the CTU label and selectively applies several two-dimensional finite impulse response filters (Wiener filters) before the image becomes output for prediction. At the encoder side, the algorithm estimates the sets of low- pass frequency-domain filter coefficients that frequently lead to minor errors and then conveys the signals to the corresponding pixel coefficients at the decoder side.

*4) Luma Mapping with Chroma Scaling:* The Luma Mapping with Chroma Scaling in-loop operation in VVC includes two basic sequential processes: 1) Luma Mapping and 2) Chroma Scaling. The Luma Mapping uses an adaptive piecewise linear reshaping function and scales the input luma code values (signal) across the dynamic ranges. The reshaping function takes the intensity lines from the input, makes it into sixteen segments on the curve, and finally boosts up the signal precision. Therefore, the luma mapping makes better use of luma code range values within specified bit depth and improves the coding efficiency for standard and high dynamic range video. On top of Luma mapping, the Chroma scaling step applies residual signaling on the Chroma components with flag activation. With the above two operations, the decoder improves the signal, adopts some signal characteristics, and thereby improves the image quality

### B. CNN based SAO Filter in Predecessor HEVC

The existing in-loop algorithms in HEVC primarily target removing the blocking artifacts. To remove the undesired complex artifacts, Park and Kim [9] proposed a deep learning-based IFCNN network as an alternative to the de-blocking filter and improve the visual quality in HEVC [1]. Kim and Lee [3] introduced a super-resolution method using very deeper residual learning and demonstrated the contextual information over a large image region. Later Jia et al. [19] proposed an iterative training scheme and introduced a content-aware deep-learning based in-loop filtering technique for quality improvement. T. Li et al. [4] proposed the MIF-Net to replace original deblocking and SAO in HEVC and investigated the image enhancement under low bit-rate intra coding configuration. Zhang et al. [24] proposed the RHCNN model to measure the mappings between the raw frames and their corresponding reconstructed frames and investigated the residual performance. Kang et al. [23] suggested a novel multi-modal sub-networks to substitute the present SAO filter at inter-intra mode. The model largely contains two CNN sub-networks with different scales and boosts the performance of the image restoration. Recently, Ma et al. [36] proposed a CNN cascaded post-processing model for multi-level feature learning, improved information flow, and shown consistent coding gains in VVC.

## III. Overall Architecture

In recent years the deep learning community brings new progress in high-level computer vision tasks and shares the idea of the deeper network for better output performance. We investigated both the traditional and heuristic learning methods on video compression and found that the heuristic-based method may not outperform the conventional inter prediction in view of reducing temporal redundancy. However, the traditional video compression framework consumes a high rate cost to signal the motion information limiting overall compression efficiency. This observation motivates us to propose a decoder-side multi-scale CNN-based inter-prediction method to efficiently obtain similar reference blocks without explicitly signaling the motion information. Considering visually annoying artifacts often observed in their reconstructed frames, a deconvolution-based filter is adopted to address this issue and improve the reconstructed video quality. Subsequently, the spatial and temporal information is jointly exploited by taking both the current block and the reference block into consideration during temporal frame-blocks processing. Our proposed model can reduce the diversity across the frames and learn the frame differences, including artifact, noise, object motion, and appearance changes.

### A. Inter Frame Operation

In VVC, the working frame is first divided into non-overlapping equal-sized CTU regions up to 128×128 samples and then further divided into smaller coding units (CU). Each CU then recursively split into four square-sized child coding blocks. A child coding block can further be split into two equal-sized children coding blocks using binary-tree (BT) split or into three unequal sized blocks using ternary-tree (TT) split or four blocks using Quad-tree split. As a consequence, the multi-type tree structure follows a flexible hierarchical partitioning tree to adapt to the content information in local image regions. A CU can be coded by using either inter or intra prediction. In intra prediction, the spatial neighboring samples are used to predict the current block. Likewise, if the inter prediction is used, one or more similar blocks will be searched from the already encoded P or B neighboring frames and predict the current block. The relative position shift between the current block and its similar blocks in the reference frame is termed as motion vector (MV) and later signaled to the decoder with motion compensation flag information. The residual block information is sent to the quantization and transform modules to generate the quantized residual coefficients, which are then encoded by using the CABAC entropy coding. After that, the residual coefficients are inverse quantized and inverse transformed to obtain the reconstructed residual block at the decoder side. The predicted residual block (intra or inter) and the current input block are added together to form the reconstructed block.

### B. Network Architecture

The block diagram of our MSCNN network is shown in Fig. 3 and later integrated locally in the decoder side of the VVC reference software VTM 8.0 [8]. The residue image is compressed in the VTM framework and feeds into the in-loop decoder network. The overall network incorporates a chain of ten convolution and symmetric deconvolution layers and jointly functions in an end-to-end framework. The convolutional filter learns the spatial information through activation maps and down-samples the input image contents into smaller size abstractions. Then the deconvolution and un-pooling layers up-sample the abstractions back to the original resolution. The intermediate convolution layers act as a feature extractor, captures the primary object components in the image through a series of activation maps, and reduces the noise components. The features extracted on the first layer (Fig. 3) are quite noisy. Therefore, we followed three immediate steps: feature de-noising, spatial feature shrinking, and feature enhancement layer for clearing the low-level attributes. The mapping layer maps these high dimension attribute vectors to another intermediate vector and restores the high-level features. Finally, the residual layer aggregates the patch-wise representation, adds the input frame patch, and generates the final output (extreme right of Fig. 3). The residual feature maps for the upper (U) and lower (L) network are defined as below:

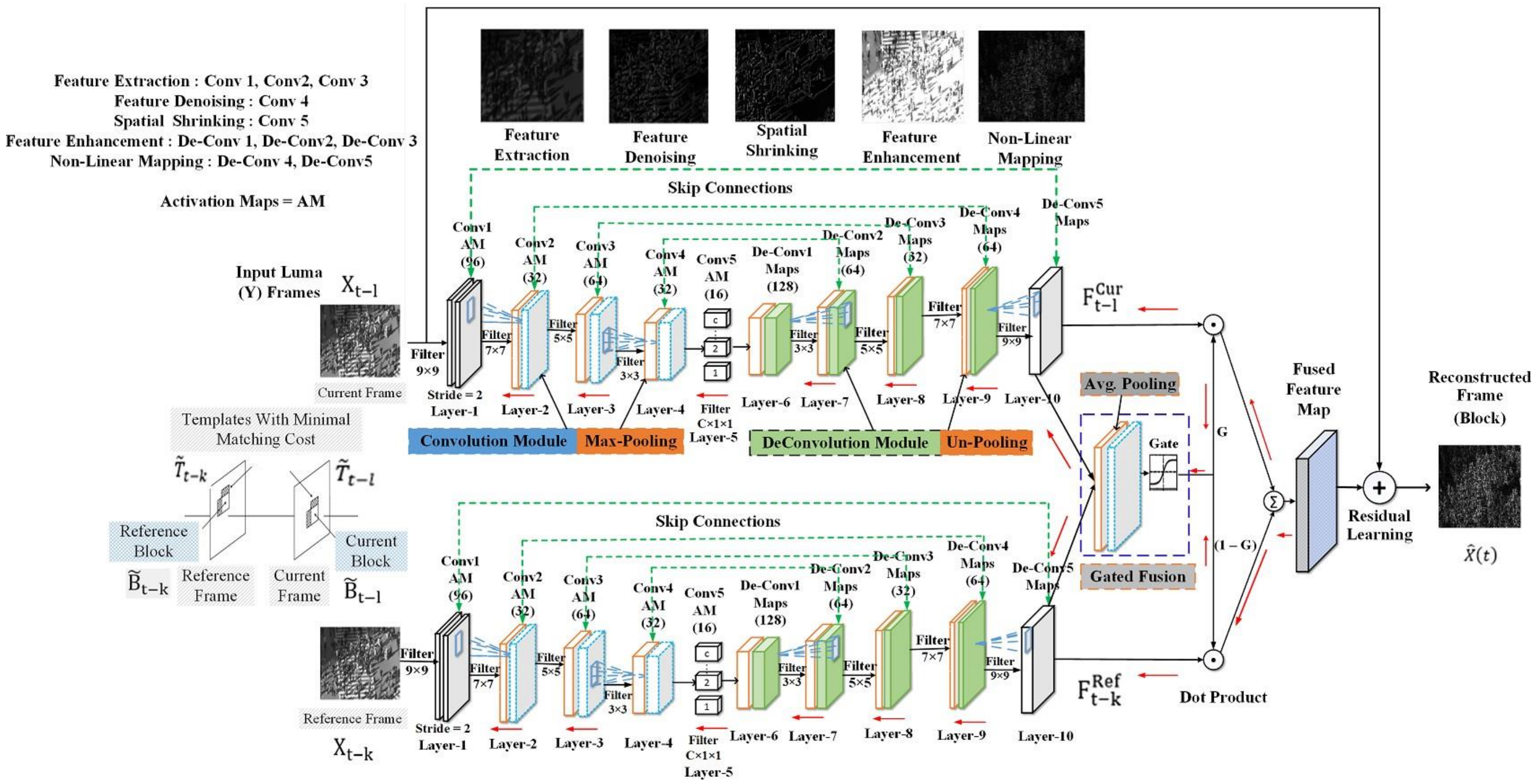

Fig. 3: Overall Architecture Diagram of our Deep Learning Model for Inter Frame Reconstruction. The Compressed Luminance (Y) input Frames from Encoder feed to the Model. Upper and Lower subdivisions are Parallel and Each Contains Symmetric Layers of Convolution, Deconvolution, and Includes: Feature Extraction, Denoising, Shrinking, Enhancement, Mapping and Reconstruction Layer. We Merged the Current and Reference Block Feature Maps to a Gate Array and weighted the Contribution of each Block for final Image Reconstruction.

$$
\begin{aligned}
F_0(Y) &= Y_0 \\
F_i(Y) &= max(0, w_i * Y_i + b_i), where\, i \in \{1 \text{ to } 10\} \\
F_{t-l}^{Cur} &= \overbrace{w_{10}^{Cur} * F_9(Y_i) + b_{10}}^{F_{10}^{Cur}(Y)} + X_{t-l} \\
F_{t-k}^{Ref} &= \overbrace{w_{10}^{Ref} * F_9(Y_i) + b_{10}}^{F_{10}^{Ref}(Y)} + X_{t-k}
\end{aligned} \tag{1}
$$

This paper proposes a novel deconvolution learning based approach for image quality enhancement and addresses the SAO filtering in VVC inter frame prediction. The shallow model learns the residual image through a gated architecture and later merges the input frames with the residue, and finally enhances the object details. We first constructed a large balanced database with different video resolutions for our model evaluation. Our deconvolution model utilizes diverse spatial features and restores the detailed information for further image quality improvement. To make the network capture more detailed information, a new loss function is included for model training. The experiment results show that our proposed network outperforms the baseline model in terms of achieving higher bit rate reduction and computational efficiency. The PSNR versus bit-rate comparison was presented and the coding efficiency of various state-of-the-art methods assessed using the same gradient-based optimization techniques. The proposed scheme was able to reduce compression artifacts while checking the rate-distortion cost after CU label reconstruction. Although our model shows a possible option for SAO filtering, more rigorous training strategies are

$th$

needed for real-time application usage. In future proceedings, we would like to include a wide-range categorical dataset and include local salient features for our experiment. In recent times the recurrent neural network is broadly used for three-dimensional video analysis [10], [37], [39]. In the future direction, we would like to explore the bidirectional LSTM models [18] for (B and P) prediction operations and extend the model to adaptive loop filtering in Versatile Video Coding. $F_i$, $w_i$, and $b_i$ in Eq. 1 represent the output feature activation map function, filter weight, bias components of $i_{th}$ layer respectively and the symbol $*$ denotes the convolution operation. The $W_i$ contains $n_i$ filters of support $n_{i-1} \times f_i \times f_i$ where $f_i$ represents the spatial size, $n_i$ equal to the number of filters, and $n_0$ as the number of channels in the input compressed image. Our layered framework is denoted as $n_1(f_1)$-$n_2(f_2)$-$n_3(f_3)$-$n_4(f_4)$-$n_5[f_5]$-$s[f_6]$-$s[f_7]$-$s[f_8]$-$s[f_9]$-$s[f_{10}]$, where f, [], n, s represents the filter size, deconvolution filters, number of convolution filters, and stride respectively. The network settings are represented as 96(9)-32(7)-64(5)-32(3)-16(1)-128[3]-64[5]-32[7]-64[9]-1[9] and described in Table I. The nonlinear mapping and spatial shrinking layers are chosen with $3 \times 3$ filters to reduce overall complexity. The features extracted from the first three layers are quite noisy. Therefore, we included $2 \times 2$ max-pool layers at every convolution layer on the encoder side and reduced the feature dimensions. A strided convolution with stride $(s)$ two is followed to speed-up the intermediate convolution operation and faster network training. In the convolution module (left in Fig. 3) we used three max pool filtering operations to reduce the spatial size and control network over-fitting. Although the pooling layer reduces the number of parameter computations, some spatial localization information is lost within a receptive field. To overcome such issues, three symmetric un-pooling layers are employed in the deconvolution module (right in Fig. 3). The un-pooling layer performs the reverse pooling operation, enlarges the activation map, and reconstructs the original size of the input. The activation maps of the output un-pooling layer are sparse. Consequently, to enlarge the receptive field and retrieve more contextual information at a less computation cost, the up-convolutions operations are incorporated at each layer. Therefore, deconvolution layers densify these sparse activations through several convoluted low pass filter operations. In this way, the deconvolutional layers associate the input activation maps with multiple output image noise locations which are often ignored in previous convolutional layers (left in Fig. 3). In our experiment, we explored both the luminance (Major) and chrominance color components and input frames to our MSCNN network. The ReLU non-linear activation function is applied to the filter output responses. The end-to-end CNN model is then designed to learn the residue F(Y) images. Therefore, the input image is combined with the residual to form the reconstructed image at the end.

TABLE I: Layer Wise Configuration Summary of MSCNN Network and Their Percentage Parameter Contribution.

| Layers | Layer 1 | Layer 2 | Layer 3 | Layer 4 | Layer 5 | Layer 6 | Layer 7 | Layer 8 | Layer 9 | Layer 10 |
|---|---|---|---|---|---|---|---|---|---|---|
| Convolution | Conv1 | Conv2 | Conv3 | Conv4 | Conv5 | Conv6 | Conv7 | Conv8 | Conv9 | Conv10 |
| Filter Size | f1: $9 \times 9$ | f2: $7 \times 7$ | f3: $5 \times 5$ | f4: $3 \times 3$ | f5: $1 \times 1$ | f6: $3 \times 3$ | f7: $5 \times 5$ | f8: $7 \times 7$ | f9: $9 \times 9$ | f10: $1 \times 1$ |
| Filters Activation Maps | n1 : 96 | n2 : 32 | n3 : 64 | n4 : 32 | n5 : 16 | n6 : 128 | n7 : 64 | n8 : 32 | n9 : 64 | n10 : 1 |
| Parameters | 23424 | 4736 | 4864 | 896 | 64 | 3584 | 4864 | 4736 | 15616 | 10 |
| Total Parameters = 70596 | | | | | | | | | | |
| Percentage(%) | 33.18 | 6.70 | 6.88 | 1.27 | 0.09 | 5.07 | 6.889 | 6.70 | 22.12 | 0.01 |

### C. Gated Feature Fusion

The gated fusion layer is introduced to effectively combine the inter-frame local and temporal features for the final image reconstruction. As illustrated in Fig. 3, the proposed fusion operation is composed of three layers: average pooling, convolution, and gated sigmoid layer. The $F_{t-l}^{Cur} \in R^{c \times h \times w}$ and $F_{t-k}^{Ref} \in R^{c \times h \times w}$ denotes the probability maps current and reference feature space at the final deconvolution layer (Fig. 3). The symbol 'c' denotes the number of feature channels or filter length, 'h' height, and 'w' the width of the map. The feature maps $F^{Cur}$ and $F^{Ref}$ from Eq. 1 are concatenated to obtain a fused probability map $F_{k-l}^{fusion} \in \mathrm{R}^{2c \times h \times w}$. Thereafter, we employed average pooling and sequentially, a spatial convolution operation with weights $W \in R^{c \times 2c \times 1 \times 1}$ and c filters with dimension of $2c \times 1 \times 1$ per filter. The weights are learned to correlate the two feature maps from different frame regions and average their contributions for the prediction of forthcoming frame blocks. Therefore, the output of the convolution layer is a coefficient matrix $G \in R^{c \times h \times w}$ and represented as:

$$G_{k,i,j} = \sum_{k'=1}^{2c} F_{k',i,j}^{fusion} \times W_{k',k,i,j} \qquad \forall k \in [1, c], i \in [1, h], j \in [1, w]. \tag{2}$$

Finally, a sigmoid squashing function is used to regularize G and map the $G_{k,i,j}$ values in range $\in [0, 1]$. We term the $G^{Cur} = G$ and $G^{Ref} = (1 - G)$ as the weighted gates, where $G_{k,i,j}^{Cur}$ and $G_{k,i,j}^{Ref}$ denote how confidently one can rely on current and reference feature maps respectively to predict the pixel (i, j) in channel k. The two coefficient matrices are then utilized to weigh the contributions of current and reference as follows:

$$\begin{aligned} \tilde{F}^{Cur} &= F^{Cur} \odot G^{Cur} \\ \tilde{F}^{Ref} &= F^{Ref} \odot G^{Ref} \end{aligned} \tag{3}$$

where $\odot$ denotes Hadamard dot product. Finally, we generated the gated fusion probability feature map as a weighted combination of $F^{Cur}$ and $F^{Ref}$.

$$\tilde{F}^{fusion} = \tilde{F}^{Cur} \odot \tilde{F}^{Ref} \tag{4}$$

We predicted the feature residue map by a $\tilde{F}()$ fusion function and leverage the current frame map to optimize the total network via stochastic gradient descent.

### D. Loss function

We investigated different model settings through a series of controlled experiments and identified a trade-off between the in-loop filtering performance and its influences on model depth. Considering the complexity of the VVC frame work, a shallow CNN model is included with depth ten. The similarity between the reconstructed and the original frame is formulated as a regression problem. To measure the model performance, we computed mean squared error from a given batch of compressed ground truth residual patches $(x_i)^n \in \{X_i\}$ and their corresponding feature maps $\tilde{F}_i^{\,n}{}^{fusion}$ through multiple epochs. To constrain the reconstruction error, we set the training loss as a combination of pixel error between the reconstructed transmission $\hat{X}_i(x)$ and the corresponding ground truth map $X_i(x)$. The total loss function is expressed as below:

$$\begin{aligned} L(\theta) &= \frac{1}{N^2} \sum_{i=1}^{n} \left\| \hat{X}_i - X_i \right\|^2 + \lambda_1 \sum_{i=1}^{l-1} \| F_i{}^l (Y_i, \theta_i) \|^2 \\ &where, F_i{}^l (Y_i, \theta) = z_j{}^i \oplus f_{i \times i}, \\ &l = no.\ of\ intermediate\ layers \end{aligned} \tag{5}$$

Where $\theta_i$ represents the network parameter and includes both weight $w_1$ to $w_{10}$ and bias terms $b_1$ to $b_{10}$, and n is the number of samples in a training batch. The $z_j{}^i$ and $f_{i \times i}$ represents $j^{th}$ layer feature with a kernel size of $i \times i$. Our goal is to learn the loss function that maps the predicted residue values from input frame compressed patches. The mapping function $F_i()$ performs the intermediate SAO filtering operation and optimizes the $\theta$ parameters through iterative training steps. This is achieved by minimizing the $L_2$ loss between the ground truth blocks $X_i$ and corresponding gated feature map $\tilde{F}^{fusion}$. The reconstructed residual images are then added to the input image to get the final reconstructed frame $\hat{X}_i$. However, the MSE loss can potentially generate blurry images and result in the loss of image details. Therefore, we included an additional loss term $F_i{}^l$ defined in the feature space $z_j{}^i$ learned from skip-connected convolution layers.

**Algorithm 1:** Optimization Procedure of the Network

**Require:** Initiate model parameters $\theta$: $(W_0, B_0)$, learning rate $(\eta)$, regularization param: $\lambda \geq 0$
**Input :** Number of iterations $N_{iter}$, epochs $N_{epoch}$. A batch of $(x_i)^n$ sampled from training set $\{X_i\}$, patch size of $128 \times 128$
**Output :** Model with mapped $\hat{X}_i$, $F_i{}^{fusion}$ and $\mathrm{F}_i^l(Y)$

**foreach** *no_epoch in* $N_{epoch}$ **do**
  **foreach** $k$ *in* $N_{iter}$ **do**
    Sample batch of 32 patches $\{x_i\}^n$
      **foreach** *i in n* **do**
        $\hat{x}_i \leftarrow x_i + F_i{}^l(\theta_i, Y_i)$ in Eq. (6)
        $\frac{\partial L_k^{(t)}(\theta_i,\lambda)}{\partial x_i} = \nabla F_i^{fusion}(\hat{x}_i, x_i, \nabla\theta_i) + 2\lambda_1 \nabla F_i{}^l(x_i, \nabla\theta_i) + \frac{\lambda}{2n}\|w_i\|^2$ in Eq. (7)
      **end**
      $\hat{X}_i = \sum_{i=1}^n \hat{x}_i$
  **end**
  Update $w_k \leftarrow w_k - \eta \frac{\partial}{\partial w_k} L(F^{fusion}(x_i, Y_i, \theta_k(w_k,|b_k)) + 2\lambda w_k$
**end**

In Algorithm 1, $x_i \in X_i$, $R^{N\times N}$ denotes the $i^{th}$ training compressed sample of a batch. The model predicts the output patch $\hat{x}_i \in Y_i$ and tries to approximate the input patch $x_i$. Our training scheme follows an iterative process with forward and backward propagation and simultaneously optimizes the model parameters. The network weight, bias, and hyper-parameter values are first initialized in Algorithm 1 and followed for $N_{iter}$ iterations. The image patches are processed per iteration and integrated at the final layer of deconvolution for full image reconstruction. To make the final output image $\hat{X}_i$ close to $X_i$ in Eq. 6, our training steps followed a mini-batch stochastic gradient descent and performed constrained optimization with the $L_2$ norm. The optimization step in Algorithm 1 includes the squared norm of the weight matrix $(\theta)$ multiplied by the regularization parameters. In Eq. 7, The regularization component drives the values of weight, bias, and $F_i(Y)$ matrix down and effectively reduces the over-fitting.

## IV. Experimental Results

To validate the performance of the model, we included a residue learning mechanism into VVC VTM 8.0 [8] reference software. The experiment results for the test sequences are shown in the result section. Our proposed MSCNN model is compared with other latest SAO filtering methods and followed the common test conditions defined in [16], [25]. To verify our algorithm performance, we validated the video test sequences of class A, B, C (WVGA), and D (WQVGA) videos. Initially, we focused on the inter coding model performance with AI mode configuration in VTM 8.0 software package. The input video I-frame sequences are encoded at various QP values {22, 27, 32, 37}. The BD-BR and BD-PSNR [12] measurements are calculated for the performance evaluation and results are shown in Table II, and III.

TABLE II: Average %BD-Rate of our Model Evaluated across the channels on AI Mode and Compared to Baseline VVC Model [8].

| Class | Sequence | Frame Count | BD-Rate (%) | | |
|---|---|---|---|---|---|
| | | | Y | U | V |
| Class A (2560 × 1600) | PeopleOnStreet | 60 | -3.93 | -4.19 | -5.53 |
| | Traffic | 60 | -4.29 | -5.45 | -4.46 |
| | SteamLocomotive | 100 | -3.24 | -3.97 | -4.32 |
| | Nebuta | 100 | -4.36 | -4.66 | -4.54 |
| Class B (1920 × 1080) | BQTerrace | 100 | -2.95 | -3.03 | -3.56 |
| | ParkScene | 100 | -3.92 | -3.47 | -4.62 |
| | BasketballDrive | 100 | -2.06 | -3.52 | -5.34 |
| | Kimono | 80 | -3.29 | -3.75 | -6.32 |
| Class C (832 × 480) | BQMall | 60 | -5.02 | -4.61 | -5.75 |
| | PartyScene | 100 | -3.43 | -3.98 | -5.23 |
| | BasketballDrill | 100 | -3.92 | -4.58 | -5.62 |
| | RaceHorses | 100 | -4.59 | -5.24 | -4.04 |
| Class D (416 × 240) | BlowingBubbles | 100 | -3.62 | -3.48 | -4.17 |
| | RaceHorses | 100 | -2.55 | -3.23 | -5.82 |
| | BasketballPass | 100 | -4.12 | -4.35 | -4.83 |
| | BQSquare | 100 | -3.46 | -4.53 | -4.95 |
| Class E (1280×720) | FourPeople | 60 | -3.92 | -5.33 | -5.38 |
| | Johnny | 60 | -4.23 | -4.37 | -5.04 |
| | KristenAndSara | 100 | -3.46 | -5.44 | -4.68 |
| Summary | Class A | | -3.955 | -4.568 | -4.713 |
| | Class B | | -3.055 | -3.443 | -4.96 |
| | Class C | | -4.24 | -4.603 | -5.16 |
| | Class D | | -3.688 | -3.898 | -4.943 |
| | Class E | | -3.87 | -5.047 | -5.033 |
| Overall | | | -3.762 | -4.312 | -4.962 |

### A. Data Pre-processing and Training Setup

For our model training, we picked seven hundred fifty input images with different resolutions from DIV2K dataset [14], eighteen video sequences of Classes A to E from the JCT- VC test set [21], and collected ninety-two supplementary sequences from Xiph.org [22]. To reduce the correlation between the channels, color images are decomposed into YUV 4:2:0 file format (chroma U and V channels half the resolution of luma Y) and encoded by VTM-8.0 at different QP values. All raw videos are encoded under All-intra, Random Access, Low-delay and Random-access configurations. The built-in inloop filters are all enabled while compressing the video sequences. The whole dataset is split into two training and testing folders and randomly choose ninety-five video sequences for our train-set (80%), and the remaining 20% kept aside for the test-set. The training and test images had no overlap to demonstrate the generalizability of model validation. During the training phase, largely the Y luma channel of video sequences have proceeded as input to the network. This is because the Y is the luminance channel and contains most of the visual information. For training the CNN model, the image pixel values of input video frames are normalized between 0 to 1 from its original intensity range. To use the training data more efficiently three image augmentation techniques are adopted: 1) rotation by degrees of 90, and 180, 2) scaling by a factor of 0.75, 0.5, and 0.25, and 3) horizontal flipping. Subsequently, our off-line augmented training set becomes twenty-four (3×4×2) times that of the original dataset.

Our MSCNN network implementation is derived on an open-source Python/C++ based Tensorflow Keras deep learning library [7]. All the experiments are carried out on a desktop I7 computer and K40 GPU@875 MHz graphic card with 12 GB memory. Each image in the training set is decomposed into n number of $128 \times 128$ overlapping patches and put into a set X = $\{X_i\}$, $i \in (1...n)$. The patch size $128 \times 128$ is chosen judiciously to fully accommodate the maximum encoder CTU structure and integrated to get a frame-wise reference after deconvolution. Each input image is compressed by intra-inter coding at four different QP's $\{22, 27, 32, 37\}$. The trained network is applied to each patch $X_i$ in an input image and subsequently constructs the de-noise output image by combining the results from all output Y = $\{Y_i\}$, $i \in (1...n)$ patches in a simple manner. The input patches are extracted from the ground truth training images with a stride of 10. Overall, the augmented $900 \times 24 = 21600$ images are converted into 4,348k training patches. The network is trained separately at each QP values and set hyper-parameters for fine-tuning. The $L(\theta)$ loss function is computed by comparing the ground-truth $X_i$ patches and the deconvoluted output image of size $128 \times 128$ pixels. For the given number of samples and computational power, our objective loss function is minimized iteratively using stochastic gradient descent with standard back propagation. The weight matrices in each layer are updated in terms of gradients and learning rates through back propagation. During the training, we used a smaller learning rate ($\eta$) of $1 \times 10^{-4}$ in the last layers and $1 \times 10^{-3}$ in the remaining layers for better convergence. The learning rate is divided by 100 when the loss value becomes stable. The momentum parameter is set to 0.9 and weight decay to 0.0001. To monitor our model performance, 20% of training data are separated for validation. In training, the reconstruction loss is used for optimization over 7240 iterations and with a batch size of 32. Different trained files are collected for subsequent QP parameters, compiled, and integrated into a locally installed VTM software [8]. For our CNN model testing, we turned off the base SAO filter from the configuration file, performed an end-to-end flow, and thereafter compared the result with the base VVC SAO performance.

## V. Result Analysis

We validated our model with classes A, B, C, D video sequences in AI mode, and their corresponding % BD-bitrate results are reported in Table II. A post-processing output comparison was performed and the results are reported in reference to the VVC SAO filter outputs. We calculated the BD-rate measures [12] on both chrominance and luminance channels and independently evaluated the coding efficiency. It was observed that our MSCNN model achieved a considerable bit-rate reduction in various test sequences listed in Table II. For the luminance (Y) frames, a high of -5.02% BD-rate is achieved on the BQMall sequences, and on an average of -3.762%, BD-rate is achieved on the rest of the sequences.

TABLE III: Average BD-PSNR and SSIM for Different Methods Across Y channels: VTM [8], MGNLF [33], and Our Model.

| Evaluation Metrics | Quantization Parameter | VTM [8] | MGNLF [33] | Our Model |
|---|---|---|---|---|
| PSNR(dB) | 37 | 26.16 | 28.68 | 31.39 |
| | 32 | 30.29 | 31.56 | 35.22 |
| | 27 | 32.25 | 32.61 | 35.74 |
| | 22 | 34.15 | 35.13 | 38.88 |
| SSIM | 37 | 0.782 | 0.824 | 0.855 |
| | 32 | 0.833 | 0.861 | 0.894 |
| | 27 | 0.902 | 0.917 | 0.923 |
| | 22 | 0.911 | 0.936 | 0.953 |

### A. Result Comparison with CNN models

We compared our MSCNN model results with VVC Baseline [8], AWT [29], and MGNLF [33]. To measure the reconstructed frame distortion, the BD-PSNR, and SSIM evaluation metrics are compared with diverse test sequences and shown in Table III. We also observed that the MSCNN network outperformed other models, based on various evaluation metrics. This indicates that our MSCNN can reconstruct the input image with less compressed artifacts. Additionally, the computational complexity of different CNN models is compared in terms of decoding time on test sequences at different resolutions (Table IV). All methods are implemented in the Keras machine learning framework and K40 16GB stand-alone GPU system.

TABLE IV: Time Complexity Comparisons of various In-Loop SAO Filters in AI mode and GPU Configuration.

| Class | Time Overhead: t {Ref}/t {VTM} in second | | | |
|---|---|---|---|---|
| | AWT [29] | VTM [8] | MGNLF [33] | Our Model |
| Class A (2560×1600) | 0.652 | 0.432 | 0.184 | 0.127 |
| Class B (1920×1080) | 0.488 | 0.516 | 0.165 | 0.155 |
| Class C (832×480) | 0.192 | 0.241 | 0.158 | 0.128 |
| Class D (416×240) | 0.184 | 0.225 | 0.242 | 0.216 |
| Class E (1280×720) | 0.285 | 0.169 | 0.465 | 0.343 |
| Average | 0.3602 | 0.3166 | 0.2428 | 0.1938 |

### B. Time Complexity

We analyzed the time complexity of our MSCNN approach and compared it with other recent in-loop filter models in GPU mode only. The decoding time overhead for each in-loop filter reference model $t_{Ref}$ is evaluated in reference to VTM baseline $t_{VTM}$ on class A to E sequences and averaged. Thereafter, the $t_{Ref}/t_{VTM}$ ratio is reported for model comparison in Table IV. A larger ratio indicates a relatively longer time overhead of the in-loop filter operation. Table IV summarizes the average run time of representative learning-based CNN methods across the entire JCT-VC test sequences with different image resolutions. The experimenters are performed in a machine learning (TensorFlow) VTM 8.0 framework with GPU mode and recorded the computational time. One can notice from Table IV that the AWT [29] based in-loop filtering approach has a time overhead among the three categories in GPU mode. However, benefiting from the GPU acceleration, our MSCNN model is faster (0.1938 sec/frame) than the AWT [29] (0.3602 sec/frame) and got a modest improvement compared to the VTM [8] performance (0.3166 sec/frame) and MGNLF [33] (0.2428 sec/frame). From the above analysis, we observe that the proposed MSCNN deconvolution network consumes less time among the rest of the network configurations and can be a holistic approach in terms of time complexity. The decoding time did not meet the required real-time application requirements and need further careful optimization.

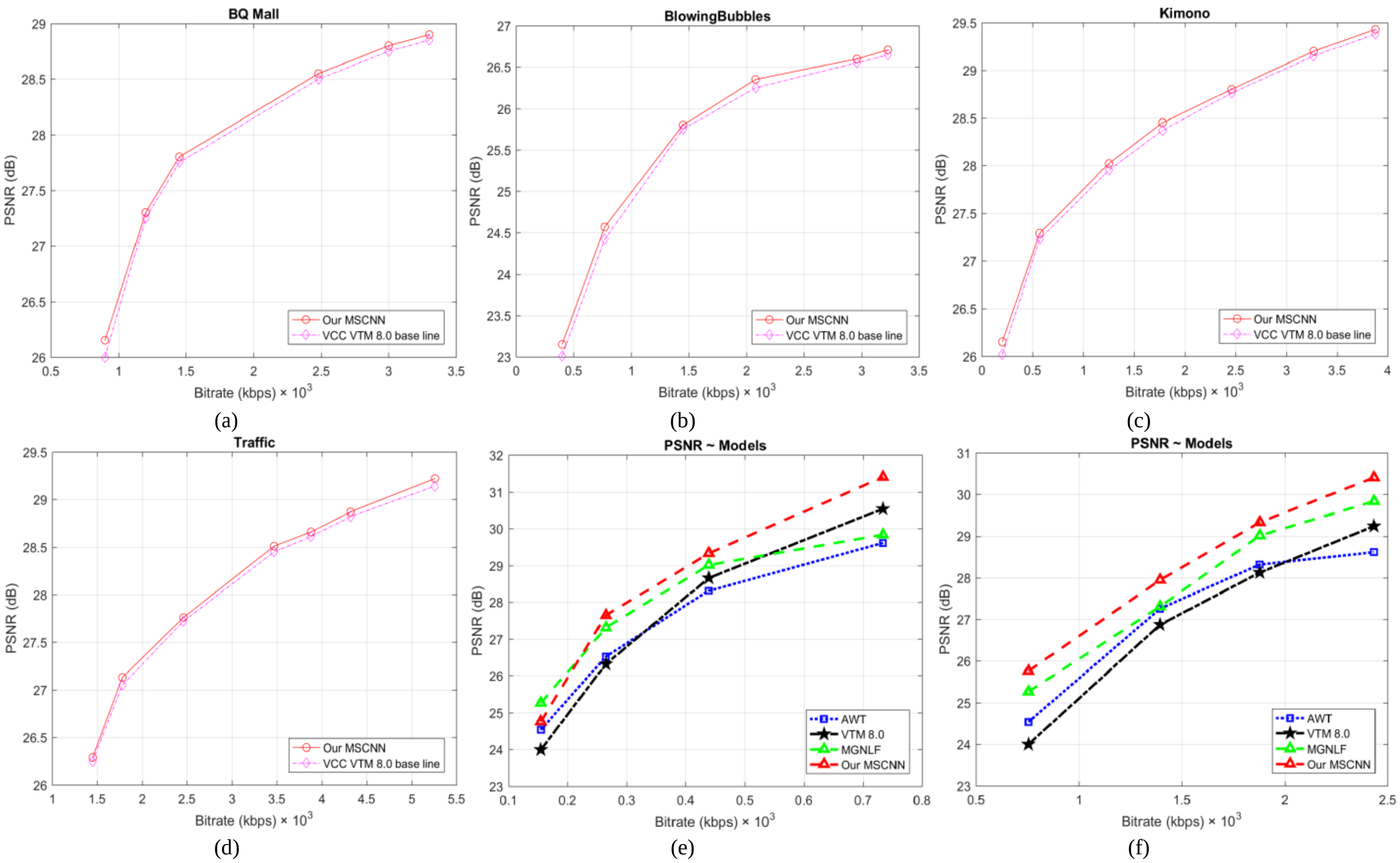

Fig. 4: Model performance comparison for different video sequences and configurations (AI/RA) in AWT [29], VTM [8], and MGNLF [33]. (a) BQMall, AI; (b) BlowingBubbles, AI; (c) Kimono, RA; (d) Traffic, AI; (e) KristenAndSara, AI mode; (f) BasketballDrill, RA mode.

### C. Objective Evaluations and Results Analysis

The quality assessment for the reconstructed video sequences in a lossy video compression model is measured together in terms of the objective and subjective performance viewpoints. The standard objective measurements are performed mathematically by using pixel differences between a compressed frame as compared to that of a reference frame. These difference values usually provide the mathematical correctness of the measurement. In our objective video analysis experiment, we estimated the structural similarity (SSIM) and peak signal-to-noise ratio (PSNR) values for different in-loop methods and compared those at different bit-rates. Table V summarizes the rate-distortion results of all three approaches in which the original VVC VTM 8.0 is used as an anchor and with baseline SAO in-loop filtering operation. The BD-BR and BD-PSNR results of our algorithm are tabulated in Table V for AI, and RA configurations, averaged and successively compared with AWT [29], VTM [8], and MGNLF [33] methods. As reported in Table V, the BD- BR of our MSCNN approach is -4.485% averaged over the conventional test sequences, and outperforming -1.943% of AWT [29] and -3.873% of MGNLF [33]. We observed that our proposed MSCNN performs better both in terms of BD- PSNR improvement and BD Bit-rate reduction. The MGNLF [33] model outperforms the AWT [29] because it involves more layers and has higher accuracy. In terms of the PSNR bit-rate match, our approach attains 1.466 dB in the standard test set and considerably well over the 0.808 dB of [29], and 1.13 dB of [33] in AI (intra I-frame), and RA configurations. Eventually, our MSCNN model achieved superior RD performance across all three approaches. The potential reasons for such maximum output include the (1) utilization of wide range activation maps, (2) proposed symmetric convolutional and deconvolutional blocks, and (3) effective feature de-noising and shrinking feature learning. The inter-frame experiment was performed in VTM software with original network settings provided in Fig. 3. Our Conv and De-Conv CNN model is separately applied to different frame blocks in a multi-scale fashion (Fig. 3), reconstructed the image features parallelly [34], and later merged in a gated fusion layer. In our inter coding experiment, each reconstructed frame are served as a reference for encoding successive P or B frames. The CNN model output is then fed back to the next successive frame in a decoded frame buffer to produce its enhanced version. The P and B frames are encoded and trained separately for respective models and followed adaptive QP training strategy addressed in LDP and LDB mode. Additionally, we performed model testing on the Xiph.org video media set at [22] and sampled thirteen test sequences from each resolution. Table VI presents the BD-BR performance of our model under AI, RA, LP, and LB configurations. The proposed architecture shows a stable performance and achieved a -4.272% bit rate saving in AI mode over the video media-set and substantially outperformed other three model configuration results as -3.63% (RA), -4.205% (LDP), and -4.134% (LDB). In terms of BD-PSNR, our LDB configuration model achieved 1.339 dB in the above test set, accomplished 0.449 dB in AI model, 1.192 dB in RA model, and 1.244 dB in LDP mode. In belief, our proposed approach performs favorably under different compression configurations and network settings. To intuitively compare the model performance, we also included the histogram statistics of PSNR values for different classes and sequences in Fig. 5. It is noted that our model achieves the highest PSNR which demonstrates the effectiveness of hierarchical feature learning to improve the compression performance.

TABLE V: BD-BR Coding Performance (bit rate), %BD-PSNR (dB) of our MSCNN and Other State-of-the-art Methods w.r.t. VTM 8.0 [8] across Y channel space, QP values, Class Resolutions and Configurations.

| Class (Resolution) | Sequence | Configuration | | | | | | | | | | | |
|---|---|---|---|---|---|---|---|---|---|---|---|---|---|
| | | AI | | | | | | RA | | | | | |
| | | AWT [29] vs VTM | | MGNLF [33] vs VTM | | Our Model vs VTM | | AWT [29] vs VTM | | MGNLF [33] vs VTM | | Our Model vs VTM | |
| | | BD PSNR | BD-BR (%) | BD PSNR | BD-BR (%) | BD PSNR | BD-BR (%) | BD PSNR | BD-BR (%) | BD PSNR | BD-BR (%) | BD PSNR | BD-BR (%) |
| Class A1 (3840×2160) | Tango2 | 1.614 | -1.102 | 2.412 | -4.23 | 1.152 | -4.18 | 0.95 | -2.52 | 1.136 | -4.293 | 1.717 | -5.06 |
| | FoodMarket4 | 1.836 | -2.163 | 1.882 | -3.57 | 1.325 | -5.41 | 1.03 | -3.07 | 1.435 | -5.018 | 1.851 | -5.35 |
| | Campfire | 1.394 | -1.881 | 1.653 | -2.35 | 1.034 | -4.13 | 0.77 | -2.37 | 1.036 | -4.016 | 1.675 | -5.14 |
| Class A2 (3840×2160) | CatRobot1 | 0.665 | -2.417 | 1.314 | -3.15 | 1.567 | -5.17 | 1.022 | -2.63 | 1.461 | -3.417 | 1.853 | -4.75 |
| | ParkRunning3 | 1.841 | -2.765 | 1.386 | -2.88 | 1.732 | -5.51 | 1.056 | -3.28 | 1.452 | -4.194 | 2.072 | -4.53 |
| | DaylightRoad | 1.475 | -2.262 | 1.514 | -2.22 | 1.652 | -4.16 | 0.813 | -2.62 | 1.162 | -4.071 | 1.883 | -5.21 |
| Class B (1920×1080) | MarketPlace | 0.916 | -3.29 | 1.035 | -3.77 | 1.483 | -4.34 | 1.057 | -2.52 | 1.585 | -4.18 | 1.819 | -4.31 |
| | BasketballDrive | 0.807 | -1.91 | 0.853 | -5.02 | 1.193 | -5.37 | 1.23 | -2.91 | 1.254 | -4.85 | 2.217 | -5.17 |
| | Cactus | 0.636 | -2.15 | 0.618 | -4.48 | 1.412 | -4.05 | 0.79 | -1.39 | 1.178 | -4.11 | 1.518 | -5.34 |
| | BQTerrace | 0.501 | -1.13 | 0.546 | -4.17 | 1.328 | -3.91 | 1.27 | -1.82 | 1.315 | -4.72 | 1.553 | -5.16 |
| | RitualDance | 0.415 | -1.08 | 0.454 | -3.64 | 1.021 | -4.58 | 0.88 | -2.19 | 0.913 | -4.57 | 1.075 | -5.12 |
| Class C (832×480) | PartyScene | 0.302 | -2.37 | 0.319 | -3.04 | 1.543 | -3.13 | 1.05 | -1.55 | 1.41 | -2.72 | 1.88 | -4.58 |
| | BasketballDrill | 0.406 | -1.61 | 0.393 | -3.74 | 1.432 | -3.88 | 0.72 | -1.34 | 1.38 | -2.44 | 1.75 | -2.93 |
| | RaceHorses | 0.435 | -2.68 | 0.481 | -4.022 | 1.425 | -4.51 | 0.54 | -1.52 | 1.84 | -3.11 | 1.64 | -4.32 |
| | BQMall | 0.511 | -2.811 | 0.442 | -4.07 | 0.822 | -4.46 | 0.49 | -1.13 | 1.42 | -3.75 | 1.78 | -5.12 |
| Class D (416×240) | BlowingBubbles | 0.862 | -2.45 | 0.796 | -4.88 | 1.281 | -3.38 | 0.692 | -2.25 | 1.089 | -3.937 | 1.81 | -3.37 |
| | BQSquare | 0.781 | -2.17 | 0.773 | -4.65 | 1.34 | -4.86 | 0.481 | -1.85 | 1.839 | -4.303 | 1.34 | -4.34 |
| | BasketballPass | 0.381 | -1.44 | 0.497 | -4.18 | 1.152 | -4.94 | 0.431 | -1.15 | 1.713 | -4.15 | 1.15 | -5.24 |
| Class E (1280×720) | KristenAndSara | 0.298 | -0.38 | 0.556 | -3.64 | 1.04 | -3.51 | 0.58 | -0.77 | 1.453 | -3.76 | 0.87 | -4.45 |
| | Johnny | 0.121 | -0.84 | 0.295 | -3.88 | 1.16 | -4.16 | 0.89 | -1.31 | 1.274 | -2.98 | 1.33 | -3.12 |
| | FourPeople | 0.354 | -1.08 | 0.735 | -4.12 | 1.35 | -3.75 | 0.647 | -1.45 | 1.183 | -4.404 | 1.36 | -4.38 |
| Overall | | 0.788 | -1.904 | 0.903 | -3.795 | 1.307 | -4.352 | 0.828 | -1.983 | 1.358 | -3.952 | 1.626 | -4.619 |

TABLE VI: The Luma Y Channel BD-BR (%) Performance of Our Model on Fifty-two Xiph.org [22] Video Test Sequences at Different Configurations and Evaluated with Respect to Baseline VVC VTM 8.0 Framework [8].

| Resolution | Sequence | | Our Model vs VTM (AI) | | Our Model vs VTM (RA) | | Our Model vs VVC (LDP) | | Our Model vs VVC (LDB) | |
|---|---|---|---|---|---|---|---|---|---|---|
| | Name | No. of | BD-BR(%) | BD-PSNR | BD-BR (%) | BD-PSNR | BD-BR (%) | BD-PSNR | BD-BR (%) | BD-PSNR |
| (1280, 720) | Mobcal | 4 | -5.412 | 0.125 | -5.382 | 1.155 | -4.652 | 1.282 | -4.146 | 1.367 |
| | Holm | 6 | -3.021 | 0.265 | -4.185 | 1.108 | -3.523 | 1.159 | -3.328 | 1.284 |
| | Shields | 3 | -5.012 | 0.157 | -3.874 | 1.122 | -2.941 | 1.211 | -4.382 | 1.312 |
| (720, 480) | Calendar | 5 | -4.271 | 0.368 | -4.584 | 1.201 | -4.573 | 1.238 | -4.536 | 1.274 |
| | Galleon | 4 | -3.785 | 0.122 | -4.095 | 1.152 | -4.028 | 1.186 | -2.182 | 1.202 |
| | WashDC | 4 | -5.618 | 2.86 | -3.036 | 1.094 | -5.325 | 1.125 | -2.951 | 1.147 |
| (640, 360) | Fountain | 5 | -2.782 | 0.205 | -3.584 | 1.316 | -4.274 | 1.285 | -5.172 | 1.294 |
| | Dance | 5 | -3.894 | 0.241 | -4.012 | 1.358 | -3.829 | 1.306 | -4.824 | 1.404 |
| | Bridge | 3 | -4.856 | 0.272 | -3.013 | 1.216 | -4.083 | 1.248 | -4.511 | 1.355 |
| (352, 288) | Bowing | 2 | -5.127 | 0.286 | -4.038 | 1.284 | -5.24 | 1.342 | -3.737 | 1.463 |
| | News | 3 | -2.023 | 0.214 | -2.804 | 1.125 | -4.027 | 1.186 | -3.844 | 1.335 |
| | Football | 6 | -5.852 | 0.348 | -4.332 | 1.244 | -4.032 | 1.316 | -4.439 | 1.427 |
| | Bus | 2 | -3.884 | 0.192 | -4.625 | 1.117 | -4.132 | 1.262 | -4.024 | 1.504 |
| Average | | 52 | -4.272 | 0.449 | -3.63 | 1.192 | -4.205 | 1.244 | -4.134 | 1.339 |

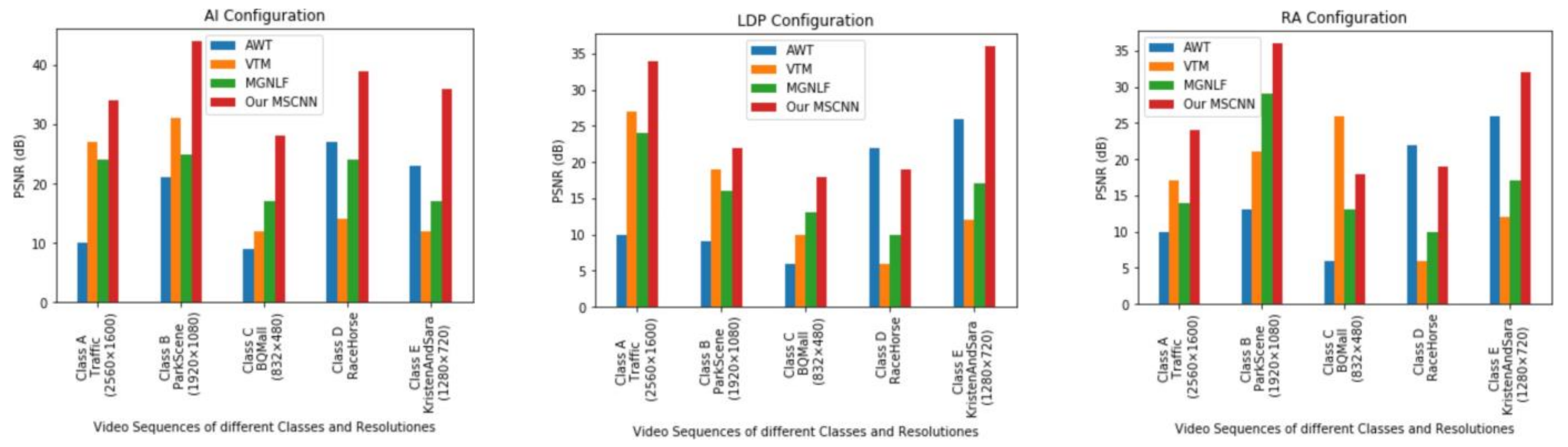

(a) AI Configuration (b) LDP Configuration (c) RA Configuration

Fig. 5: PSNR histogram statistics of AWT [29], VTM [8], MGNLF [33] models and configurations on different test video sequences.

Fig. 4 a) to d) shows the Rate-Distortion (RD) curves and illustrates the bit-rate savings of our model. The figure additionally shows the PSNR performance comparisons of the proposed MSCNN network and VVC SAO de-blocking filters in luminance frames on four selected sequences: BQ Mall, BlowingBubbles, Kimono, and Traffic sequences. To further understand the performance, another set of experiments are performed at AI and RA configuration mode in KristenAndSara and BasketballPass sequences (Fig. 4 e and f). We used a constant quality of ten bits per pixel to represent each pixel in the frame and measured the number of bits for encoding the representations. The optimal bit rate in $kb/s \times 10^3$ is calculated by multiplying the above bits per pixel and video resolution. The graphical comparison between PSNR~Bitrate (kbps) is shown in Fig. 4 e) and f). It can be viewed that the PSNR gain of our MSCNN method is higher than that of AWT [29], VTM [8] methods and a slight edge in performance to that of MGNLF [33] at different QP values. As shown in Fig. 4 e), the AWT model output looks inconsistent at higher QP values. The main reason could be, it failed to predict high-frequency wavelet features at different filter sizes and uses more memory at the decoder buffer. In brief, the proposed method achieved better feature generalization in our test dataset and obtained consistent evaluation performance under different configurations.

## D. Subjective Results

In this section, we analyzed the variation of decoded frame quality which serves as a premise for our proposed model. Subsequently, subjective measurements are performed and evaluated the perceived visual image quality that the human eye may not notice. Fig. 6 exhibits visual image quality comparisons of the original in-loop SAO filter in VVC [8], MGNLF [33], AWT [29], our MSCNN approach, and demonstrate the effectiveness of various approaches. The input "BQMall" sequence has a resolution 832×480 and the residual transfer block coefficients are quantized at a factor of $QP32$. For the quality comparison, we used the fifth frame of "BQMall" video and reconstructed the images. We put a bounding box area around the single person of a zoomed-in region of "BQMall" (right) for small object reconstruction comparison. Although the traditional HEVC filters have been applied in Fig. 6 a), one can notice that there is still blurring, and the visually ringing effect persists in the baseline results. In Fig. 6 b), the frames filtered by the VVC model look flattened, and a sudden intensity variation is observed through the image objects. Subsequently, one can observe that the reconstructed frames in Fig. 6 c) in the third quadrant (left) are over-smoothed especially in the leg, bag, and shawl region, and introduce noisy lines on the shallow boundary regions. Moreover, some details are missing in the crowded region of the bottom-left corner and minor color distortion at the top right of the elliptical edge regions. On the contrary, Fig. 6 (d) refers to the results of employing the proposed MSCNN model. Our proposed adaptive model successfully recovered the feature details around the bag edge, hand and collar regions with less compression artifact. After post-processing, the smooth details are preserved to a greater extent and look visually appealing. This perhaps is due to the better generalization ability of our model training. However, some details are missing at the shop wall corner and leg of glasses.

TABLE VII: Ablation Study on Different Model Component Combinations. Computational Time, BD-BR, and BD-PSNR Performance Evaluated and Reported on SJTU Video Sequences [28].

| No | Multi-Scale Network | BD-PSNR (dB) | BD-BR (%) | Time (in Sec) |
|---|---|---|---|---|
| 1 | Conv5 + Padding + Up-Sampling | 1.148 | -4.52 | 0.374 |
| 2 | Conv5 + Un-pooling + De-Conv1 | 1.436 | -5.46 | 0.388 |
| 3 | Conv5 + Un-Pooling + De-Conv2 | 1.191 | -4.61 | 0.457 |
| 4 | Conv5 + Un-Pooling + De-Conv3 | 1.525 | -4.84 | 0.484 |
| 5 | Conv5 + Un-Pooling + De-Conv4 | 1.365 | -5.07 | 0.518 |
| 6 | Conv5 + Un-Pooling + De-Conv5 | 0.942 | -5.34 | 0.454 |
| 7 | Conv5 + Average Up-Pooling + De-Conv5 | 1.502 | -4.08 | 0.747 |
| 8 | Conv5 + Un-Pooling + De-Conv5 + Modes | 1.278 | -4.28 | 0.385 |

## E. Ablation Study

In this section, we conducted a series of ablation studies and validated the efficacy of our model components. Our

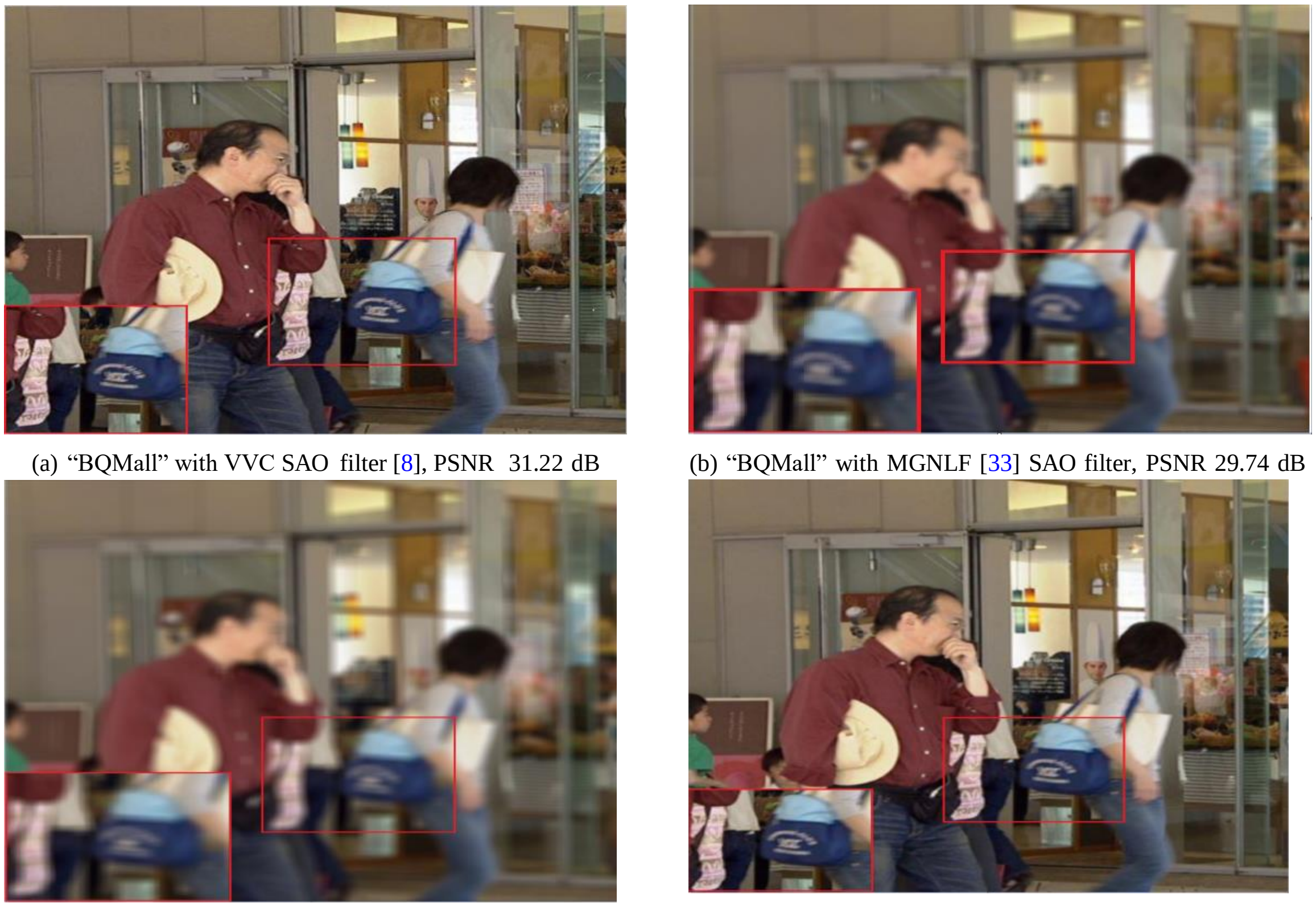

(a) "BQMall" with VVC SAO filter [8], PSNR 31.22 dB

(b) "BQMall" with MGNLF [33] SAO filter, PSNR 29.74 dB

(c) "BQMall" with AWT [29] SAO filter, PSNR 30.35 dB

(d) Our MSCNN Model, PSNR 32.07 dB

Fig. 6: Visual quality of a compressed "BQMall" with RA configuration and $832 \times 480$ resolution. The fifth frame with QP = 32 is shown at different SAO filter models.

analysis starts from the CNN convolution module and then incrementally replaced or removed components to understand the effectiveness of SAO filter operation. To verify the image enhancement, we kept the model convolution module intact and varied the deconvolution and pooling layers in both upper and lower network of our CNN model (Fig. 3). Consequently, we came up with eight networks with different components and listed them in the second column of Table VII. For a fair comparison, the testing procedures for each ablation study are kept exactly the same. We report the results on SJTU test Sequences [28] in Table VII. In network "7", we replaced the un-pooling operation with the un-average pooling. Our conventional average pooling layer computes the mean value of the activation maps within a filter window. From the results, one can found that more coding gain achieved with network "7", but at the cost of the decoder computation time. Although the average pooling operation achieved robust feature representation in noise, it introduced the blurring effect in object boundaries. One can note that the base Network "8" with all given model components achieves the best image enchantment performance in terms of BD-BR, and BD-PSNR evaluation. One can notice that the Conv5 activation maps embed the rich features and put across those cues into deconvolution network. The top Conv4 and Conv5 layers have more contextual information and combining deconvolution operation on these feature map benefits the overall filtering operation. The Network "4", "5", and "6" suggest that the up-convolution feature learning plays a pivotal role in noise reduction and possibly because of the propagation of fine-grained features during our model learning.

## VI. CONCLUSION

This paper proposes a novel deconvolution learning-based approach for image quality enhancement and addresses the SAO filtering in VVC inter frame prediction. The shallow model learns the residual image through a gated architecture and later merges the input frames with the residue, and finally enhances the object details. We first constructed a large balanced database with different video resolutions for our model evaluation. Our deconvolution model utilizes diverse spatial features and restores the detailed information for further image quality improvement. To make the network capture more detailed information, a new loss function is included for model training. The experiment results show that our proposed network outperforms the baseline model in terms of achieving higher bit rate reduction and computational efficiency. The PSNR versus bit-rate comparison was presented and the coding efficiency of various state-of-the-art methods assessed using the same gradient-based optimization techniques. The proposed scheme was able to reduce compression artifacts while checking the rate-distortion cost after CU label reconstruction. Although our model shows a possible option for SAO filtering, more rigorous training strategies are needed for real-time application usage. In future proceedings, we would like to include a wide-range categorical dataset and include local salient features for our experiment. In recent times the recurrent neural network is broadly used for three-dimensional video analysis [10]. In the future direction, we would like to explore the bidirectional LSTM models [18] for (B and P) prediction operations and extend the model to adaptive loop filtering in Versatile Video Coding.